\documentclass[12pt,a4paper,final]{iopart}

\usepackage[utf8]{inputenc}

\expandafter\let\csname equation*\endcsname=\relax 
\expandafter\let\csname endequation*\endcsname=\relax 
\usepackage{graphicx, bbold, amssymb}
\usepackage{color}
\usepackage{mathtools}
\usepackage{hyperref}

\usepackage{dcolumn}

\begin{document}

\newcommand{\fd}[1]{{\color{blue}{#1}}}
\def\vec#1{\bm{#1}}
\def\ket#1{|#1\rangle}
\def\bra#1{\langle#1|}
\def\braket#1#2{\langle#1|#2\rangle}
\def\ketbra#1#2{|#1\rangle\langle#2|}

\title{Reservoir engineering of Cooper-pair-assisted transport with cold atoms}

\author{Fran\c{c}ois Damanet$^{1}$, Eduardo Mascarenhas$^1$, David Pekker$^{2,3}$, and Andrew J.\ Daley$^{1}$}

\address{$^{1}$ Department of Physics and SUPA, University of Strathclyde, Glasgow G4 0NG, United Kingdom.\\
$^{2}$ Department of Physics and Astronomy, University of Pittsburgh, Pittsburgh, Pennsylvania 15260, USA.
\\
$^{3}$ Pittsburgh Quantum Institute, Pittsburgh, Pennsylvania 15260, USA.
}

\begin{abstract}

We show how Cooper-pair-assisted transport, which describes the stimulated transport of electrons in the presence of Cooper-pairs, can be engineered and controlled with cold atoms, in regimes that are difficult to access for condensed matter systems. 
Our model is a channel connecting two cold atomic gases, and the mechanism to generate such a transport relies on the coupling of the channel to a molecular BEC, with diatomic molecules of fermionic atoms. Our results are obtained using a Floquet-Redfield master equation that accounts for an exact treatment of the interaction between atoms in the channel. We explore, in particular, the impact of the coupling to the BEC and the interaction between atoms in the junction on its transport properties, revealing non-trivial dependence of the produced particle current. We also study the effects of finite temperatures of the reservoirs and the robustness of the current against additional dissipation acting on the junction. Our work is experimentally relevant and has potential applications to dissipation engineering of transport with cold atoms, studies of thermoelectric effects, quantum heat engines, or Floquet Majorana fermions.  

\end{abstract}


\section{Introduction}

Transport measurements between reservoirs connected by a channel are well-known tools to understand and study the static and dynamical properties of condensed matter systems. In this context, the development of cold atom platforms has offered possibilities to explore phenomena with strongly-interacting particles in transport setups. A key feature of these setups is that they can be described by microscopic models derived from first principles under well-controlled approximations~\cite{Kri17}.  Such setups allow for the simulation of novel phenomena and exploration of the fundamental mechanisms since they allow for tuning of the microscopic parameters such as interaction and potential. Examples include the observation of quantised transport of neutral atoms in a junction connecting cold gas reservoirs~\cite{Kri15}, or the investigation of the role of interaction and temperature on transport in quantum point contacts~\cite{Hus15} or lattices~\cite{Leb18}. 

In addition to connecting to well-known phenomena of solid state physics, cold atom platforms offer the possibility to investigate new paradigms of transport, via continuous measurements~\cite{Uch18} or dissipation engineering. Indeed, the atomic motion occurs on sufficiently long timescales that the transient dynamics can be measured and controlled in real time. These tools have been long applied in few-body systems in quantum optics~\cite{Wis09, Gar15}, and in that context form the basis for standard techniques such as laser cooling and trapping~\cite{Met99}. The coupling to reservoirs is well-understood microscopically under well-controlled approximations, and can be engineered experimentally. In the transport channel, particle losses, which naturally occur via collisions with a background gas, can be engineered via the use of an electron beam~\cite{Ger08} or light scattering through a quantum gas microscope with single-site resolution~\cite{Wei11,Bak09,She10}. Dephasing can also be realized via light scattering or noise sources~\cite{Dal14,Lus17,Sar14,Pic10,Nie18}.

Taking advantage of the level of microscopic control offered by cold atoms, we study here transport of fermionic atoms between reservoirs weakly connected by a single site junction, a system that resembles a quantum dot junction connecting leads. In particular, we explore the possibility to control transport based on Andreev reflection, i.e., transport of electrons assisted by exchange of Cooper-pairs~\cite{And64, Bee97, Mar11}, via reservoir engineering. In contrast with~\cite{Hus15}, the junction we consider is a weakly-connected single site -- not a quantum point contact -- and our reservoirs are non-interacting. Such a system naturally produces sequential tunnelling of atoms, yielding a quantised particle current. We show here how to engineer the transport between the reservoirs by coupling the junction to a molecular BEC~\cite{Jia11}, mimicking Cooper-pair assisted transport of electrons in the solid-state, and yielding a rich peak structure in the current-bias characteristics. We then study the effects of finite temperature of the reservoirs and interaction between atoms in the junction on the produced current, and also determine its robustness against the effects of particle losses acting on the channel.

Our results are obtained using a Floquet-Born-Markov (or Floquet-Redfield) master equation~\cite{Yan16,Koh97,Gra94, Gri98}, which goes beyond standard Gorini-Kossakowski-Sudarshan-Lindblad (GKSL) master equations~\cite{Lin76, Gor76, Bre06}. Such a method, in addition to treat the interaction in the junction exactly, makes it possible to capture the complex interplay between driving and dissipation mechanisms, and has been applied in the context of photon-assisted transport or Landau-Zener tunnelling~\cite{Bla15}. In~\cite{PhysRevLett.123.180402}, we adapted the Floquet master equation formalism to Cooper-pair driving, which appears in quantum dot systems coupled to superconductors, and demonstrated control of Cooper-pair-assisted transport of electrons. Here, we use this framework in a cold atom context, where the driving comes from the molecular BEC and the dissipation processes correspond to coupling of atoms into and out off large (thermal) reservoirs.

The results we present here demonstrate the possibility to engineer transport based on Andreev reflection in cold atoms in an unconventional setup -- without the need for interactions in the source and drain reservoirs -- and in regimes that are hard to access with other methods. Our work also provides a framework to diagnose the impact on transport of many effects that could be engineer experimentally, such as controlled interaction and dissipation.
We also analyse realistic experimental conditions, including finite temperatures in the reservoirs.

The paper is organised as follow. In Sec.\ II, we detail our model and summarise the main steps of the derivation of the master equation used to calculate the transport properties of the junction. In Sec.\ III, we present our results for the particle current, with and without coupling with the molecular BEC. We study the effects of finite temperature of reservoirs, interaction between atoms and particle losses acting on the junction. In Sec.\ IV, we summarise and provide an outlook. We use in the remainder of this paper natural units in which $\hbar = k_B = 1$.

\section{Model}

\begin{figure}[tb]
\begin{center}
\includegraphics[width=0.9\textwidth]{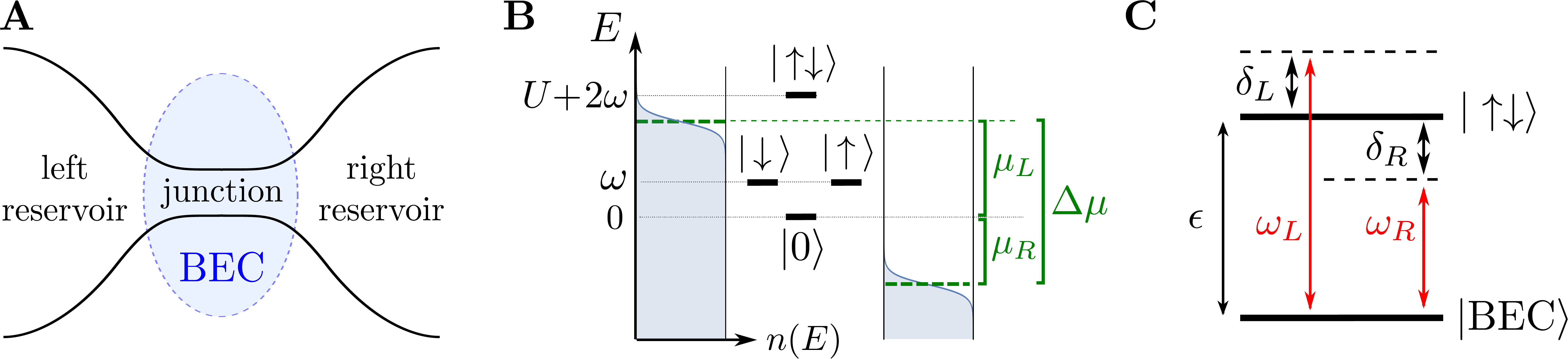}
\end{center}
\caption{\textbf{A}: Two ultracold fermionic gases connected together by a junction immersed into a molecular BEC. \textbf{B}: Energy diagram of the bare junction and occupation $n(E)$ of the reservoirs as a function of chemical potential bias $\Delta\mu$. \textbf{C}: Atom-molecular conversions in the junction induced by two fields of radiofrequencies $\omega_L$ and $\omega_R$ and detunings $\delta_L = \omega_L - \epsilon  > 0$ and $\delta_R = \omega_R - \epsilon < 0$, where $\epsilon$ is the frequency of the transition $|\mathrm{BEC}\rangle \leftrightarrow  |{\uparrow\downarrow}\rangle = c^\dagger_\uparrow c^\dagger_\downarrow | 0\rangle $, inspired from~\cite{Jia11}.}
\label{setup}
\end{figure}

In this section, we summarise our model for a tunnel junction connecting two cold atom reservoirs. Figure~\ref{setup} (A) shows a setup where two ultracold fermionic gases are connected together by a small junction. We consider two different spin states, labelled with $s \in \{\downarrow,\uparrow \}$. Transport of atoms through the junction is generated by preparing an initial chemical potential imbalance between the two reservoirs. We propose here to control the transport properties of the junction by immersing it into a molecular BEC and coupling them via radiofrequency fields, as explained below.

\subsection{Hamiltonian}

We consider a ``system-bath'' decomposition where the ``system'' corresponds to the junction and the ``bath'' to the cold atom reservoirs. This is described by the total Hamiltonian $H_\mathrm{tot} = H_\mathrm{S}^\mathrm{eff}(t) + H_\mathrm{B} + H_\mathrm{I}$ with
\begin{align}\label{HS}
&H_\mathrm{S}^\mathrm{eff}(t) = H_\mathrm{S} + H_\mathrm{BEC}(t) = \omega \sum_{s = \downarrow,\uparrow} c_s^\dagger c_s + U c^\dagger_\uparrow c_\uparrow c_\downarrow^\dagger c_\downarrow + \sum_{\ell} \left( g_\ell e^{i \delta_\ell t}  c_\downarrow c_\uparrow  + \mathrm{h.c.}\right), \\\label{HB}
&H_\mathrm{B} = \sum_{\ell = L,R} \sum_{k,s = \downarrow,\uparrow} \omega_{k}b_{\ell ks}^\dagger b_{\ell ks}, \\
\label{HI}
&H_\mathrm{I} = \sum_{\ell = L,R} \kappa_\ell \sum_{k, s = \downarrow,\uparrow} \left(  b_{\ell ks}^\dagger c_s  + \mathrm{h.c.} \right),
\end{align}
where $H_\mathrm{S}(t)$ is the effective Hamiltonian of the junction including the influence of the molecular BEC, $H_\mathrm{B}$ the sum of the Hamiltonian of the left (L) and right (R) reservoirs, and $H_\mathrm{I}$ the tunnelling Hamiltonian between the junction and both reservoirs, where $\kappa_\ell$ is the tunnelling amplitude of atoms between the reservoir $\ell$  ($\ell = L,R$) and the junction.

In Eq.~(\ref{HS}), $H_\mathrm{S} =  \omega \sum_{s} c_s^\dagger c_s + U c^\dagger_\uparrow c_\uparrow c_\downarrow^\dagger c_\downarrow$ corresponds to a single-site Hubbard model (Anderson impurity model) for fermionic atoms of energy $\omega$, spin $s \in \{\downarrow,\uparrow \}$, and interaction $U$. We consider for the sake of simplicity the same energy $\omega$ for both spin $s \in \{\downarrow,\uparrow \}$, even though such assumption can be relaxed without any difficulty. Hence, the bare junction is an effective system of dimension $d_S = 4$ spanned by the non-occupied, single occupied, and double-occupied states $\left\{\ket{0},\ket{{\downarrow}},\ket{{\uparrow}},\ket{{\downarrow\uparrow}}\right\}$. 
The corresponding potential geometry could be achieved as proposed in~\cite{Bru12} by using two laser beams with adjusted detunings, beam waists, and positions, but also more generally with acousto-optical deflectors~\cite{Lin90} or holographic mask techniques~\cite{Pas08}. The interaction $U$ between atoms in the junction can for its part be tuned locally via optically-induced Feshbach resonances~\cite{Aru19}.

The last term of Eq.~(\ref{HS}), $H_\mathrm{BEC}(t) = \sum_{\ell} ( g_\ell e^{i \delta_\ell t}  c_\downarrow c_\uparrow  + \mathrm{h.c.})$, describes the effects of the coupling of the fermions of the junction to the background molecular BEC~\cite{Joc03, Gre04, Zwi03, Hol01}. Such coupling could be realized using one~\cite{Jia11} or multiple fields $\ell$ of radio-frequencies $\omega_\ell$ and detunings $\delta_\ell = \omega_\ell - \epsilon$, where $\epsilon$ is the frequency related to the transition between the molecular BEC and the pair states, i.e., $|\mathrm{BEC}\rangle \leftrightarrow |{\uparrow\downarrow}\rangle$ illustrated in Fig.~1 (C). Note that we work in the rotating-frame associated to $\epsilon$, absorbed in the definition of $\omega$ to not burden the notations. The coupling strength $g_\ell = \langle S \rangle \Omega_\ell$ of each field is determined by the macroscopic ground state occupation $\langle S \rangle$ of the BEC and the Rabi frequency $\Omega_\ell$, which can be tuned independently through different field amplitudes. It turns out that the Hamiltonian $H_\mathrm{BEC}(t)$ well-represents the so-called proximity effects induced by $s$-wave superconductors of chemical potentials $\delta_\ell/2$ and Cooper-pair tunnelling amplitudes $g_\ell$, when their superconducting gap is larger than the junction frequency scales~\cite{Mar11, PhysRevLett.123.180402}. For this reason, we consider in the following only two driving fields whose detunings are adjusted to the chemical potential of the fermionic reservoirs, i.e.,
\begin{equation}
\delta_\ell \equiv 2\mu_\ell \quad\quad  \ell = L,R,
\end{equation}
even though, in principle, any frequencies could be chosen. This choice is motivated to resemble the case of a quantum dot tunnelling junction connecting two superconducting leads, where the Cooper-pair condensates have energies related to an applied bias voltage.

In Eq.~(\ref{HB}), $b_{\ell k s}$ is the annihilation operator of a fermion of energy $\omega_k$, spin $s$ and momentum $k$ in the $\ell$ reservoir ($\ell = L,R$). We consider both reservoirs initially prepared in thermal states $\rho_\ell$ defined as
\begin{equation}\label{thermal}
\rho_\ell  = \frac{e^{- \beta_\ell \left(H_\mathrm{B} - \mu_\ell N_\ell\right)}}{\mathrm{Tr}\left[e^{- \beta_\ell \left(H_\mathrm{B} - \mu_\ell N_\ell\right)}\right]},
\end{equation}
with chemical potential $\mu_\ell$, temperature $T_\ell = 1/(k_B \beta_\ell)$, and where $N_\ell = \sum_{ks} b_{\ell k s}^\dagger b_{\ell k s}$. Various techniques have been realised to implement initial imbalance between atomic reservoirs, as summarised in~\cite{Kri17}.

\subsection{Master equation for the driven junction}

We treat the coupling of the driven junction with the left and right reservoirs in the weak-coupling regime. This justifies our ``system+bath'' decomposition and motivates the use of an open system approach. As in~\cite{PhysRevLett.123.180402}, we derive a Floquet-Redfield master equation, i.e., a Redfield master equation for the periodic time-dependent system~\cite{Yan16,Koh97,Gra94, Gri98} -- which corresponds to the driven junction in our case. In contrast with~\cite{PhysRevLett.123.180402} where the reservoirs were in a gapped phase, we consider them in a normal, non-interacting phase. This allows us to show that Cooper-pair-assisted transport can be achieved between the drain and source reservoirs even if these latter do not contain any pairs. We present below the key assumptions of the derivation of the master equation (all details can be found in Appendix A).

\subsubsection{Born and Markov approximations.}

The first key approximation is the Born approximation, which supposes that the total system-bath density matrix $\rho_\mathrm{tot}^I(t)$ in interaction picture with respect to $H_0(t) \equiv  H_\mathrm{S}^\mathrm{eff}(t) + H_\mathrm{B}$ can be written in the separable form 
\begin{equation}\label{Bornapp0}
\rho_\mathrm{tot}^I(t) \approx \rho^I(t) \otimes \rho_L \otimes \rho_R,
\end{equation}
where $\rho^I$ is density matrix of the driven junction in interaction picture and where $\rho_\ell$ are the thermal states given by Eq.~(\ref{thermal}). This approximation amounts in considering that the initial states of the bath are sufficient to determine the whole evolution of the system during a timescale $\tau_R \propto 1/\kappa_\ell^2$, the typical time scale needed for the junction to reach a non-equilibrium steady state. Note that this present model cannot describe the complete relaxation of the whole ``junction+reservoir'' system towards a common equilibrium, which occurs on a time scale larger than $\tau_R$~\cite{CommentNJP2019001}.

Using the ansatz~(\ref{Bornapp0}) and tracing over the bath degrees of freedom, the equation for $\rho^I(t)$ in second-order in $H_\mathrm{I}(t)$ reads
\begin{equation}\label{EQ000}
\begin{aligned}
&\dot{\rho}(t) = \\
&-\sum_{\ell, s}\int_0^t \bigg\{ \left[ 
 \langle B_{\ell s}^\dagger(t) B_{\ell s}(t-t') \rangle_B \left( c_s(t) c_{s}^\dagger(t-t') \rho(t - t')  - c_s^\dagger(t-t') \rho(t-t') c_s(t) \right) + \mathrm{h.c.} \right] \bigg. \\ 
 &\hspace{0.2cm}+ \bigg. \left[ 
 \langle B_{\ell s}(t) B_{\ell s}^\dagger(t-t') \rangle_B \left( c_s^\dagger(t) c_{s}(t-t') \rho(t-t')  - c_s(t-t') \rho(t-t') c_s^\dagger(t) \right) + \mathrm{h.c.} \right]  \bigg\} dt',
\end{aligned}
\end{equation}
where we removed the superscript $^I$ to simplify the notation, and where $\langle B_{\ell s}^\dagger (t_1) B_{\ell s}(t_2)\rangle_B \equiv \mathrm{Tr}_B\left(B_{\ell s}^\dagger(t_1)B_{\ell s}(t_2)\rho_\ell\right)$ is the bath correlation function with $B_{\ell s}(t) = \kappa_\ell \sum_{k}  e^{-i \omega_k t} b_{\ell ks}$. The Markov approximation consists of setting $\rho(t-t')\approx \rho(t)$ and extending the upper limit of integration to infinity. This amounts to neglect the memory effects, in the sense that this transforms the integro-differential equation into a time-local differential equation.

Both the Born and Markov approximations are justified for $\tau_R \gg \tau_B$, where $\tau_B$ is the decay time of the bath correlation function $\langle B_{\ell s}^\dagger(t_1) B_{\ell s}(t_2)\rangle$. Such a condition can in general be satisfied in different ways. In the present case, this is mainly due to the large size of the bath, which yields an infinite summation over destructively-interfering modes in the expression of the bath correlation function, making them decaying quickly (see~\ref{tauB} for a detailed analysis of these approximations)~\cite{C2, Dam19a, Tim08}.

\subsubsection{Final form of the master equation}

To obtain the final form of the master equation from~(\ref{EQ000}), we need to evaluate the time-dependence of the system operators $c_s(t) =  U(t)^\dagger c_s U(t)$ where $U(t)= \mathcal{T} e^{- i \int_0^{t} H_\mathrm{S}^\mathrm{eff}(t') dt'}$ is the system propagator with $\mathcal{T}$ the time-ordering operator, before performing the time-integration. We use for this purpose the Floquet theory, assuming that the driving is periodic of period $T = 2\pi/\Delta\mu$, where $\Delta\mu = \mu_L - \mu_R$ corresponds to the chemical potential bias between the reservoirs. We consider for simplicity $\mu_L = - \mu_R= \Delta\mu/2$. All details are given in Appendix B. The resulting master equation for the density matrix elements $\rho^{ab}(t) = \langle \phi_a(t) | \rho(t) |\phi_b(t) \rangle$ in the basis of the periodic Floquet modes $\ket{\phi_a(t)} = \ket{\phi_a(t+T)}$ labelled by indices $a = 1,\cdots,d_S$ reads, in the Schrödinger picture,
\begin{equation}\label{EQF0}
\dot{\rho}^{ab}(t) = -i(E_a - E_b) \rho^{ab}(t) + \sum_\ell (\mathcal{L}_\ell[\rho(t)])^{ab},
\end{equation}
where $E_a$ are the quasienergies corresponding to the Floquet modes $\ket{\phi_{a}(t)}$ and
\begin{equation}\label{Liouvillian}
\begin{aligned}
&(\mathcal{L}_\ell[\rho(t)])^{ab} \\
&= -\sum_{s}\sum_{k,k'\in\mathbb{Z}}\sum_{c,d}\left\{ \left[ e^{i (k + k') \Delta\mu t}
 \left( c_s^{ack} c_s^{\dagger cdk'}  \Gamma_{\ell +}(-\Delta_{cdk'})  +
 c_s^{\dagger ack} c_s^{cdk'} \Gamma_{\ell -}(-\Delta_{cdk'})\right) \rho^{db}(t) \right. \right. \\
&\hspace{1cm}\phantom{\times\bigg[}
\left. \left. - e^{i (k + k') \Delta\mu t} \left(  c_s^{ack} c_s^{\dagger dbk'}  \Gamma_{\ell -}(-\Delta_{ack}) + c_s^{\dagger ack} c_s^{dbk'} \Gamma_{\ell +}(-\Delta_{ack})\right) \rho^{cd}(t) \right] + \mathrm{h.c.} \right\}
\end{aligned}
\end{equation}
is the Liouvillian. This latter is written in terms of Fourier components $c_s^{ack} = \frac{1}{T} \int_0^T e^{-i k \Delta\mu t} \bra{\phi_{a}(t)}c_s\ket{\phi_b(t)} dt$ and complex rates
\begin{equation}\label{complexratesMT}
\begin{aligned}
\Gamma_{\ell\pm}(E) &= \int_0^\infty dt' f_{\ell \pm}(t') e^{i E t'} = \gamma_{\ell \pm}(E) + i \Omega_{\ell \pm}(E)
\end{aligned}
\end{equation}
evaluated at energies $\Delta_{abk} = E_a - E_b + k \Delta\mu$  ($k \in \mathbb{Z}$) with\begin{equation}\label{ratesMT}
\begin{aligned}
&\gamma_{\ell \pm}(E) = \gamma_\ell  [1-n_\ell(E \pm \mu_\ell)], \\
&\Omega_{\ell \pm}(E) =  \frac{\gamma_\ell}{\pi} \,\mathrm{P.V.} \int_{-\infty}^{\infty} d\omega \frac{n_\ell(\omega)}{E + \omega \pm \mu_\ell},
\end{aligned}
\end{equation}
where $\gamma_\ell = \pi \kappa_\ell^2 \rho_{\ell,N}$ is the tunnelling rate between the junction and the reservoir $\ell$ with $\rho_{\ell,N}$ its density of states assumed to be constant over the relevant frequency range,
where $n_\ell(E) = 1/(1+e^{\beta_\ell E})$ is the Fermi distribution, and where $\mathrm{P.V.\ }$ denotes the principal value. Hence, while for a standard Redfield master equation (i.e., without the driving) the rates~(\ref{complexratesMT}) are evaluated at transition between bare system energies, our Floquet-Redfield theory captures transition between quasienergies of the driven system up to multiple of $\Delta\mu$. This quantity corresponds to the energy difference obtained from the conversion of a molecule into a pair via the field of detuning $\mu_L$ which is then reconverted into a molecule via the other field of detuning $2\mu_R$, i.e., the process [see Fig.~1 (C)]
\begin{equation}\label{conversion}
|\mathrm{BEC}\rangle \xrightarrow[2\mu_L]{} |{\uparrow\downarrow}\rangle  \xrightarrow[2\mu_R]{}  |\mathrm{BEC}\rangle.
\end{equation} 
Hence, our theory describes the assisted transport of atoms thanks to the energy provided by molecular conversions. In solid-state systems, such assistance would correspond to transfers of Cooper-pairs between superconductors: currents based on multiple Andreev reflections. 

\section{Transport properties}

Solving the master equation~(\ref{EQF0}) allows us to compute the transport properties of the driven junction. We focus here on the steady state current of atoms leaving the junction to reach the right reservoir, which is defined as
\begin{equation}\label{IRdef}
\langle I_\mathrm{R} \rangle = - \sum_{s = \uparrow, \downarrow} \mathrm{Tr}\left[ c_s^\dagger c_s \mathcal{L}_R \left[  \rho_\mathrm{SS}\right] \right],
\end{equation}
where $\mathcal{L}_R[\cdot]$ is the Liouvillian~(\ref{Liouvillian}) for the right reservoir and $\rho_\mathrm{SS}$ the steady state density matrix (see Appendix B for details and expressions of other currents, such as the current of molecules in the BEC). We investigate below the ``current-voltage" characteristics of the junction, where the voltage corresponds to the chemical potential bias $\Delta\mu$. For the sake of simplicity, we consider in the reminder of this paper identical left and right tunnelling rates $\gamma_L = \gamma_R = \gamma$ and reservoir temperatures $T_L = T_R = T$.

\subsection{Without coupling to the molecular BEC}

When the junction is not coupled to the molecular BEC ($g_\ell = 0$ $\forall \ell = L,R$), the system Hamiltonian is time-independent and the Floquet-Redfield master equation reduces to a standard Redfield master equation that can be solved analytically. The steady state current in the right reservoir reads
\begin{equation}
\label{IRg0}
\langle I_{R}\rangle = 4 \gamma  \frac{n_R(\omega - \mu_R) \left(n_L(U+\omega-\mu_L)-1\right)-n_L(\omega - \mu_L) \left(n_R(U+\omega-\mu_R)-1\right)}{n_L(\omega - \mu_L)-n_L(U+\omega-\mu_L)+n_R(\omega - \mu_R)-n_R(U+\omega-\mu_R)+2}.
\end{equation}
We focus in the following on the particle-hole symmetric case, for which the double occupied state $\ket{{\downarrow\uparrow}}$ of the junction has the same energy than the non-occupied state $\ket{0}$, i.e., when $U + 2 \omega = 0$. This simplifies the analysis -- giving rise to a single parameter $U$ to characterise the bare junction energy -- and corresponds to the situation where the driving of the transition $\ket{0} \leftrightarrow \ket{{\downarrow \uparrow}}$ gives maximal effects. In that case, the current~(\ref{IRg0}) becomes 
\begin{equation}\label{IRg0}
\langle I_{R}\rangle = 2\gamma\frac{ \sinh\left(\frac{\Delta\mu}{2T}\right)}{\cosh\left(\frac{U}{2T}\right) + \cosh\left(\frac{\Delta\mu}{2T}\right)},
\end{equation}
which corresponds to a smooth step function. For $2T \ll \Delta\mu$, the quantization of the current becomes more obvious, since we have
\begin{equation}\label{IRg00}
\langle I_{R} \rangle \approx 2\gamma\, \frac{1}{e^{\frac{|U| - \Delta\mu}{2T}} + 1},
\end{equation}  
where we see that for $|(|U| - \Delta\mu)| \gg 2T$, the current goes to $0$ at small bias $\Delta\mu < |U|$ and to $2\gamma$ at large bias $\Delta\mu > |U|$. 

\subsection{With coupling to the molecular BEC}

Coupling the atoms to the BEC drastically changes the transport properties of the junction, since the sequential tunnelling of atoms can in that case be assisted by molecular conversions. 
Figure 2 (A) shows the steady state current of atoms reaching the right reservoir for different coupling $g_\ell \equiv g$ -- taken identical for both RF fields --, fixed value of (attractive) interaction $U <0$, and zero temperature. Peaks of currents appear at chemical potential bias 
\begin{equation}
\Delta\mu = \frac{|U|}{2k+1}, \quad\quad k \in \mathbb{N},
\end{equation}
as can be obtained from the resonant condition
\begin{equation}\label{RC}
\mu_L + k \Delta\mu  = \frac{|U|}{2},
\end{equation}
where $|U|/2$ is the energy of the transitions $\ket{0} \leftrightarrow \ket{s}$ and $\ket{s} \leftrightarrow  \ket{{\downarrow\uparrow}}$. Equation~(\ref{RC}) means that the maximal energy of an incoming atom (from the left reservoir) combined with multiple of the energy provided by the molecular conversion process~(\ref{conversion}) must be at least equal to the junction transition energy to generate transport. This explains why a non-zero current appears for lower bias values compared to the uncoupled case $g = 0$ [see dashed black line in Fig.~2 (A), corresponding to Eq.(\ref{IRg0})]. These peaks can be interpreted as transport based on Andreev reflections of order $k$, where the energy of $k$ Cooper-pairs are required to generate transport. Increasing $g$ cranks up the amplitude of the Andreev peaks.

Finite temperature of the reservoirs smears out the peaks, as can be seen in Fig.~2 (B). For low bias, the current decreases as a polynomial as a function of the chemical potential bias. For moderate temperature, signatures of Andreev transport can still be observed. However, For large temperature, thermal effects dominate and the effect of the driving becomes indistinct.

\begin{figure}[tb]
\begin{center}
\includegraphics[width=0.95\textwidth]{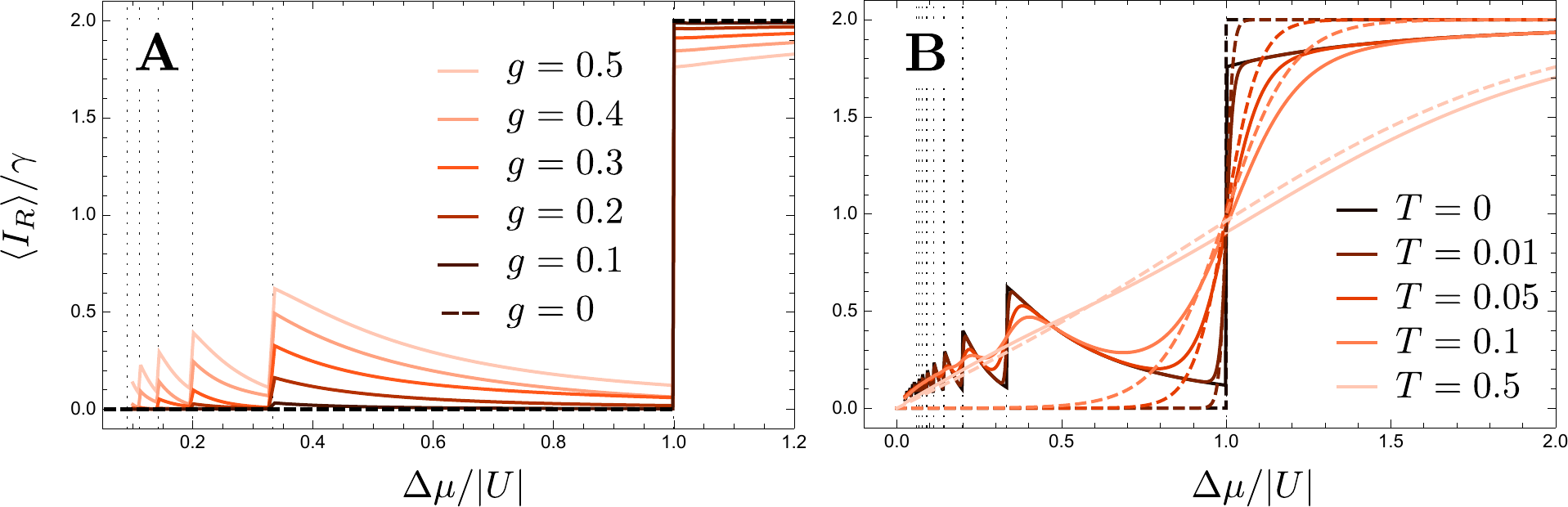}
\end{center}
\caption{Current-bias characteristics of the junction for different driving amplitudes $g$ at zero temperature $T = 0$ (\textbf{A}) and for different temperatures $T$ for fixed driving amplitude $g = 0.5$ (solid lines) and $g = 0$ (dashed lines) (\textbf{B}). Other parameters are $\omega = -U/2$ and $U = -2$, in units chosen so that $\gamma = 10^{-2}$. \textbf{A}: For $g = 0$, the current exhibits a step at $\Delta\mu = |U|$ (dashed black line). When $g$ increases, current peaks appear at $\Delta\mu = |U|/(2k+1)$ with $k \in \mathbb{N}$. \textbf{B}: Increasing the temperature smears out the peaks. In addition, for large temperature, the differences between the cases with and without driving fade.}
\label{fig2}
\end{figure} 

\subsection{Effects of interaction $U$}

In this section, we analyse the effects of the interaction $U$ in the channel on the produced transport. Figure 3 shows the current~(\ref{IRdef}) for a fixed value of $g$, zero temperature $T = 0$, and different $U$, still focusing on the particle-hole symmetric case by adjusting $\omega = -U/2$ so that we always have $U + 2 \omega = 0$. In order to compare the curves appropriately, we rescaled the chemical potential bias $\Delta\mu$ of each curve by $|U|$, which makes the peaks overlap. Otherwise, a smaller interaction $|U|$ requires a smaller chemical bias $\Delta\mu$ to generate transport. Such methodology allows us to compare the shape and the size of the Andreev current peaks for different interaction strengths. 

The system exhibits two different regimes of transport. For  $|U|/3 \leq g$ [Fig.~3 (A)], the current is characterised by small oscillations, whose the period and amplitude increase for increasing $|U|$. These oscillations are fragmented in sections 
\begin{equation}
\Delta\mu \in \left[\frac{|U|}{2(k+1)+1}, \frac{|U|}{2k+1} \right]
\end{equation}
separating the different order $k$ of multiple Andreev reflections. Around $|U|/3 \sim g$, the oscillations are no more visible and leave the place to well-resolved peaks. The amplitudes of the peaks are maximum in this regime. For $|U|/3 > g$ [Fig.~3 (B)], the amplitudes of the peaks decrease as $|U|$ increases. We thus recover the fact that Andreev reflection is suppressed for large interaction $U$. However, while it is commonly assumed that interaction has always a detrimental effects on current based on Andreev reflection in quantum dot junction~\cite{Mar11}, it seems there exists an optimal value of $|U|$, i.e., $|U|/3 \sim g$, for observing large and well-resolved current peaks. We confirmed this behaviour by considering different values of $g$ (not shown)~\cite{C3}.

\begin{figure}[tb]
\begin{center}
\includegraphics[width=0.95\textwidth]{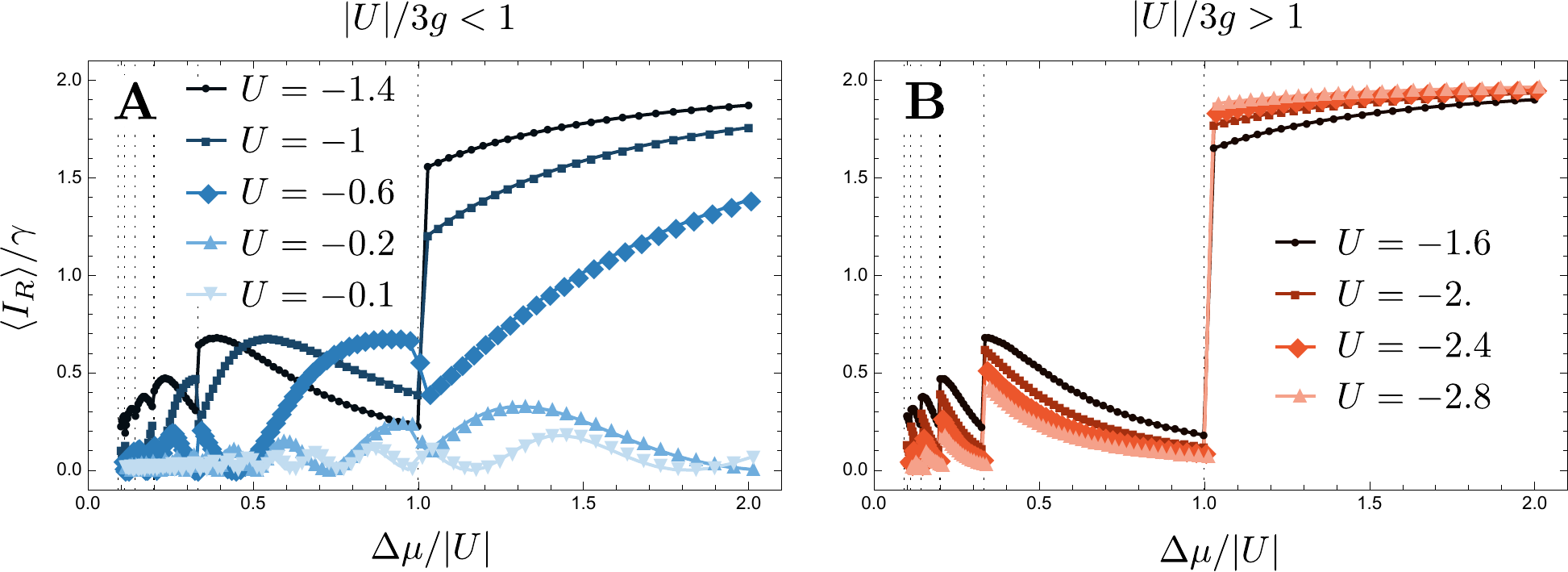}
\end{center}
\caption{Current-bias characteristics of the junction for different interaction strengths in the regime $|U|/3g < 1$ (\textbf{A}) and $|U|/3g > 1$ (\textbf{B}). Other parameters are $g = 0.5$, and $T = 0$, in units chosen so that $\gamma = 10^{-3}$. \textbf{A}: When the driving dominates, the currents is characterized by fragmented oscillations. \textbf{B}: By contrast, when the interaction starts to dominate, clear Andreev peaks appear. However, for strong interactions, i.e., $|U|/3g \gg 1 $, Andreev transport is suppressed and the current goes to its value without driving [i.e., Eq.~(\ref{IRg0})].}
\label{fig3}
\end{figure} 

\subsection{Effects of particle losses in the channel}

We finally investigate the effects of the presence of additional particle losses acting on the junction. A diagnostic of such effects is important, since particle losses are inherent in experiment due to light scattering or collisions with other atoms. This is also important to identify potential interesting consequences on transport, since particle losses can also be engineered intentionally. The main goal here is to determine whether the engineered current is robust against dissipation or not.

\begin{figure}[tb]
\begin{center}
\includegraphics[width=0.95\textwidth]{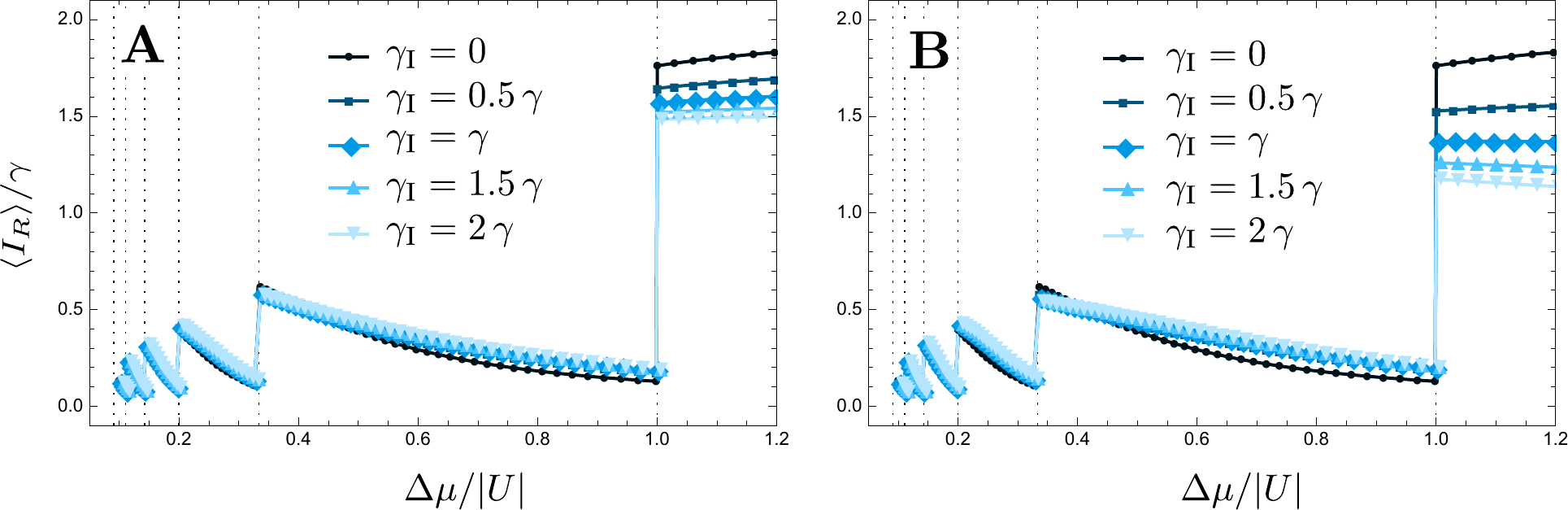}
\end{center}
\caption{Current-bias characteristics of the junction for different particle loss rates $\gamma_\mathrm{I}$ of spin down ($s = \downarrow$) atom only (\textbf{A}) and both spin up ($ s = \uparrow$) and down ($s = \downarrow$) atoms (\textbf{B}). Other parameters are $\omega = -U/2$, $U = -2$, $T = 0$, in units chosen so that $\gamma = 10^{-2}$. In both cases, the current assisted by molecular conversion is relatively robust against atom losses.}
\label{figdissip}
\end{figure}

We incorporate these effects into our master equation through an additional dissipator of the Lindblad form $\mathcal{D}_\mathrm{I}(\rho) = \gamma_\mathrm{I} \left(2 L \rho L^\dagger - \left\{ L^\dagger L, \rho \right\}\right)$, where $\gamma_\mathrm{I}$ is the rate of the incoherent process and $L$ the corresponding Lindblad operator (see Appendix C). Such dissipator corresponds to the effect of a structureless bath, but one could easily investigate the effects of a more complex bath following the procedure we used to calculate dissipation with the reservoirs. We consider in the following atom losses, where $L = c_s$ ($s \in \{ \downarrow, \uparrow \}$). Figure~\ref{figdissip} shows the atomic current in the right lead as a function of the bias potential for different loss rates $\gamma_\mathrm{I}$ of only one of the atomic species (A) and of both atomic species (B), i.e., with one dissipator of the form above for each $s \in \{\downarrow,\uparrow\}$.
For $\Delta\mu > |U|$, we observe a decrease of the current of atoms reaching the drain reservoir, since some of the atoms are lost in the additional decay channel. Surprisingly, the current assisted by molecular conversions (for $\Delta\mu < |U|$) seems to be only slightly affected by the losses, even for loss rate $\gamma_\mathrm{I}$ of the order of magnitude of the tunnelling rate $\gamma$ with the reservoirs. This is because the dissipation processes coming from the additional particle losses do not account for the driving. Hence, while the standard tunnelling processes (for $\Delta\mu > |U|$) are significantly altered by atom losses, the Cooper-pair assisted current seems resilient against them, even at higher order (i.e., at lower bias).

\section{Proposed experimental implementation}

Here we address the details of the proposed experimental implementation and observation of the phenomena we discuss in this manuscript. We give examples of parameters currently available in experiments, basing values on $\mathrm{^6}$Li atoms following~\cite{Jia11, Kri17, Aru19, Leb18}, and retain $\hbar$ and $k_B$ in our expressions within this section.

The largest frequency scale of our model is related to the transverse trapping frequency $\omega_\perp$ of the atoms, as required to avoid to take into account higher transverse modes. This transverse frequency is of the order of few dozen of $\mathrm{kHz}$ in typical experiments, but can in principle be increased using higher laser power.

Let us first consider the required temperatures. Our results show that resolved current peaks appear for $k_B T \ll \hbar \omega$. Knowing that experiments regularly reach temperatures $T \sim 50\, \mathrm{nK}$, this would require a junction frequency $\omega \sim 2\pi\times 10\,\mathrm{kHz}$ or higher, while making sure that $\omega_\perp > \omega$. 
The coupling between the states of atoms on the junction site and molecules in the molecular BEC can be strong, i.e., we can have $g_\ell \sim \omega$, as s-wave pairing as large as $2\pi\hbar \times 25\,\mathrm{kHz}$ is achievable for a BEC density of $n_0 = 10^{14}\,\mathrm{cm}^{-3}$ and molecular scattering length of $0.1\,\mathrm{nm}$~\cite{Jia11}. In addition, the locality of the coupling can be ensured by a combination of the geometry in the Raman beams, and the tuning of the near-resonant coupling between the level of the chemical potential in the molecular BEC reservoir and the two-particle trap state on the junction site.

Finally, the smallest parameter of our model $\gamma_\ell$ should be smaller than $\omega$ to satisfy the Born-Markov approximation. Indeed, observing transport at the energy $\hbar \omega$ requires a bias $\Delta\mu$, which determines the decay of the bath correlation function (see \ref{tauB}). This would mean $\gamma_\ell \sim 2\pi\times 1\, \mathrm{kHz}$ according to the other parameters above.

All together, we thus require ideally
\begin{equation}
\hbar \gamma_\ell, k_B T \ll \hbar \omega \sim \hbar g_\ell \ll \hbar \omega_\perp,
\end{equation}
and note that the transport dynamics we look at occurs on a timescale of few inverse tunnelling rates $\gamma_\ell^{-1} \sim \mathrm{ms}$. Hence, in order to resolve the current peaks as presented above, detrimental effects due to imperfections such as heating or atom losses should be small on this timescale~\cite{ComLoss}. If this is not the case, a possible solution consists in increasing $\gamma_\ell$, which implies increasing the junction and transverse frequencies $\omega$ and $\omega_\perp$. We finally note that tuning the interaction $U$ locally inside the junction using optically-induced Feshbach resonances as realised in~\cite{Aru19} provides a good route to control over interactions with strongly reduced spontaneous scattering of photons, in a form that would be appropriate for the proposed setup.

\section{Conclusion}

We showed how transport of fermionic atoms through a junction connecting two cold gases can be assisted by molecular conversion with a BEC. We described such reservoir engineering using an open-system framework that we recently derived, which is able to capture the effects of finite temperature of the reservoirs, strong interaction and presence of additional dissipation in the junction. This allowed us to explore with cold atoms the physics of Andreev reflection -- a well-known paradigm in condensed-matter -- in new parameter regimes. As a main result, we showed that there exists an optimal range of interaction yielding well-resolved, maximal peaks of assisted particle current. We showed that increasing the temperature of the reservoirs smears out the peaks, whereas these latter are robust against additional (Lindblad) particle losses acting on the junction. 

Our framework describes naturally dissipative processes and could be generalized to include the effects of measurements~\cite{Uch18} and feedback loops, to potentially engineer and uncover new phenomena in quantum transport. In addition, it could be used to study spin-polarised~\cite{Leb19} or thermoelectric~\cite{Gre16, Bra13, Sek16} transport properties of an engineered junction, starting from initial spin or temperature imbalances. Finally, since our method is suited to describe the interplay between driving and dissipation, it could be applied in the context of quantum heat engine or Floquet Majorana fermions~\cite{Ben15, LiY14, Jia11}.

\ack{FD would like to thank Konrad Viebahn for helpful discussions. Work at the University of Strathclyde was supported by the EPSRC Programme Grant DesOEQ (EP/P009565/1), and by the EOARD via AFOSR grant number FA9550-18-1-0064. Work at the university of Pittsburgh was supported by NSF PIRE-1743717. }

\appendix
\addcontentsline{toc}{section}{Appendices}

\section{Derivation of the master equation}

Our starting point is the usual Liouville-Von Neumann equation~\cite{Bre06}
\begin{equation}\label{LVN}
\dot{\rho}_\mathrm{tot}^I(t) = - i\left[ H_\mathrm{I}^I(t),\rho_\mathrm{tot}^I(t)\right] 
\end{equation}
for the total density matrix $\rho_\mathrm{tot}^I(t)$ in interaction picture with respect to $H_0(t) \equiv  H_\mathrm{S}^\mathrm{eff}(t) + H_\mathrm{B}$, where $
 H_I^I(t) = \sum_\ell \kappa_\ell \sum_{k s} (b_{\ell ks}^{I\dagger}(t) c_s^I(t) +  \mathrm{h.c.} )$
with
\begin{align}\label{ctbt}
&c_s^I(t) =  U(t)^\dagger c_s U(t),\\
&b_{\ell ks}^I(t) = e^{i H_\mathrm{B} t} b_{\ell ks} e^{-i H_\mathrm{B} t} = e^{-i \omega_k t} b_{\ell ks},
\end{align} 
where the propagator $U(t)$ is defined as
\begin{equation}\label{propa}
U(t)= \mathcal{T} e^{- i \int_0^{t} H_\mathrm{S}^\mathrm{eff}(t') dt'},
\end{equation}
with $\mathcal{T}$ the time-ordering operator. 

\subsection{Born and Markov approximations} 

Under the Born approximation, the total system-bath density matrix $\rho_\mathrm{tot}^I(t)$ can be written in the separable form 
\begin{equation}\label{Bornapp}
\rho_\mathrm{tot}^I(t) \approx \rho^I(t) \otimes \rho_L \otimes \rho_R,
\end{equation}
where $\rho^I$ is the junction density matrix in interaction picture and where $\rho_\ell$ are the thermal states given by Eq.~(\ref{thermal}). Expanding Eq.~(\ref{LVN}) up to the second order in $H_\mathrm{I}$, using the Born approximation~(\ref{Bornapp}) and tracing over the bath degrees of freedom yields
\begin{equation}\label{ME0}
\begin{aligned}
&\dot{\rho}^I(t) = -\int_0^{t} dt' \mathrm{Tr_B}\left(\left[ H_\mathrm{I}^I(t), \left[H_\mathrm{I}^I(t-t'),\rho^I(t-t') \otimes \rho_L \otimes \rho_R \right] \right]\right),
\end{aligned}
\end{equation} 
where we neglected the first order term~\cite{Bre06}. After performing the Markov approximation by setting $\rho^I(t-\tau) \approx \rho^I(t)$ and extending the upper limit of integration to infinity, the expansion of the double commutator yields
\begin{equation}\label{EQ00}
\begin{aligned}
&\dot{\rho}(t) = \\
&
- \sum_{\ell}\sum_{s} \int_0^t \bigg\{ \left[ 
 \langle B_{\ell s}^\dagger(t) B_{\ell s}(t-t') \rangle_B \left( c_s(t) c_{s}^\dagger(t-t') \rho(t)  - c_s^\dagger(t-t') \rho(t) c_s(t) \right) + \mathrm{h.c.} \right] \bigg. \\ 
 &\hspace{2.cm}+ \bigg. \left[ 
 \langle B_{\ell s}(t) B_{\ell s}^\dagger(t-t') \rangle_B \left( c_s^\dagger(t) c_{s}(t-t') \rho(t)  - c_s(t-t') \rho(t) c_s^\dagger(t) \right) + \mathrm{h.c.} \right]  \bigg\} dt'
\end{aligned}
\end{equation}
where we removed the superscript $^I$ to not burden the notation and where $\langle B_{\ell s}^\dagger (t_1) B_{\ell s}(t_2)\rangle_B \equiv \mathrm{Tr}_B\left(B_{\ell s}^\dagger(t_1)B_{\ell s}(t_2)\rho_\ell\right)$ are the bath correlations with $B_{\ell s}(t) = \kappa_\ell \sum_{k} b_{\ell ks}(t)$. Since for reservoirs in thermal states we have
\begin{equation}\label{nth}
\begin{aligned}
&\langle b_{\ell ks}^\dagger b_{\ell  k^\prime s^\prime} \rangle_B = \delta_{kk'}\delta_{ss'} n_\ell(\omega_k - \mu_\ell), \\
&\langle b_{\ell  ks} b_{\ell k^\prime s^\prime}^\dagger \rangle_B = \delta_{kk'}\delta_{ss'} (1 - n_\ell(\omega_k - \mu_\ell)), \\
&\langle b_{\ell ks} b_{\ell k^\prime s^\prime} \rangle_B = \langle b_{\ell  ks}^\dagger b_{\ell k^\prime s^\prime}^\dagger \rangle_B = 0,
\end{aligned}
\end{equation}
where $n_\ell(E) = 1/(1+e^{\beta_\ell E})$ is the Fermi occupation number, 
the bath correlation functions can be rewritten as
\begin{equation}\label{fpm}
\begin{aligned}
f_{\ell +}(t') &= \langle B_{\ell s}^\dagger(t) B_{\ell s}(t-t') \rangle_B =  \kappa_\ell^2 \sum_k e^{i \omega_k t'} n_\ell(\omega_k -  \mu_\ell)  = \frac{\gamma_\ell}{\pi} \int_{-\infty}^{\infty} d\omega \, e^{i \omega t'} n_\ell(\omega-\mu_\ell)\\
f_{\ell -}(t') &= \langle B_{\ell s}(t) B_{\ell s}^\dagger(t-t') \rangle_B =  \kappa_\ell^2 \sum_k e^{-i \omega_k t'} \left[1 -  n_\ell(\omega_k -  \mu_\ell)\right] \\
& \hspace{4cm} = \frac{\gamma_\ell}{\pi} \int_{-\infty}^{\infty} d\omega \, e^{-i \omega t'} \left[ 1 - n_\ell(\omega-\mu_\ell)\right],
\end{aligned}
\end{equation}
where $\gamma_\ell = \pi \kappa_\ell^2 \rho_{\ell,N}$ is the tunnelling rate between the junction and the reservoir $\ell$ defined below Eq.~(\ref{ratesMT}).

\subsection{Justification of the Born-Markov approximation}
\label{tauB}

As mentioned in the main text, the Born-Markov approximation is usually justified when $\tau_B \ll \tau_R \propto \gamma_\ell^{-1}$, where $\tau_B$ is the decay time of the correlation function $f_{\ell +}(\tau)$ given in Eq.~(\ref{fpm}). Here, we give an estimate of $\tau_B$.
\\
Let us first discuss the case of zero temperature $T \equiv T_\ell = 0$, where the Fermi distribution corresponds to the Heaviside distribution. For infinite chemical potential $\mu_\ell \to \infty$, the integration in $f_{\ell +}(\tau)$ involves all frequencies that interfere with each other to yield  $f_{\ell +}(\tau) = 2 \gamma_\ell\delta(\tau)$, showing that the reservoir acts as a pure Markovian reservoir with vanishing $\tau_B$. For finite $\mu_\ell$, we have however
\begin{equation}
f_{\ell +}(\tau) = 2 \gamma_\ell \delta(\tau) - \frac{\gamma_\ell}{\pi}\frac{\sin(\mu_\ell\tau/2)}{\tau/2} e^{i \mu_\ell \tau/2},
\end{equation}
where the last term decays as a power law $\sim 1/\tau$. We can thus define a typical decay time $\tau_B \propto 1/\mu_\ell$, as in~\cite{Tim08}, so that the condition of the Born-Markov approximation reads $\gamma_\ell \ll \mu_\ell$. Hence, considering a system reservoir coupling smaller than the chemical potential should in principle be sufficient to neglect non-Markovian effects. 
\\
For finite $T$ and $\mu_\ell = 0$, we have by contour integration
\begin{equation}
f_{\ell +}(\tau) = -i \gamma_\ell \frac{T}{\sinh(\pi T \tau)},
\end{equation}
for which we can define the decay time $\tau_B \propto 1/T$. Finally, for finite $T$ and $\mu_\ell$, we have
\begin{equation}
f_{\ell +}(\tau) = -\frac{\gamma_\ell}{\pi} T e^{(i \mu_\ell - \pi T)\tau} B_{-e^{\mu_\ell/T}}(1- i T \tau, 0),
\end{equation}
where $B_z(a,b)$ is the incomplete beta function, and one can estimate $\tau_B$ as the minimum of $1/T$ and $1/\mu_\ell$. Hence, we see that the Born-Markov should be satisfied for either $T \gg \gamma_\ell$ or $\mu_\ell \gg \gamma_\ell$. Note that this simple discussion does not take into account the fact that, in practice, reservoirs have a finite bandwidth, which also modifies the natural lifetime of the bath correlation function.

\subsection{Quasi-energies and Floquet states}

In order to perform the time-integration in Eq.~(\ref{EQ00}), we now evaluate the time dependence of the system operators $c_s^I(t)$ given by Eq.~(\ref{ctbt}) using the Floquet theory~\cite{Yan16,Koh97,Gra94, Gri98}. For that purpose, we suppose in the following that $\mu_L = -\mu_R = \Delta\mu/2$, so that effective system Hamiltonian $H_\mathrm{S}^\mathrm{eff}(t)$ is periodic of period $T = 2\pi/\Delta\mu$. If it was not the case, one could simply work in the rotating-frame with respect to one of the driving frequency $2\mu_\ell$, let say $2\mu_L$. This would provide a periodic Hamiltonian of period $\delta = 2(\mu_R - \mu_L)$, and the same theory would apply.

Since $H_\mathrm{S}^\mathrm{eff}(t)$ is periodic, the system wavefunction $\ket{\psi(t)}$ satisfying the Schr\"odinger equation
\begin{equation}\label{TDse}
i \frac{d}{dt}\ket{\psi(t)} = H_\mathrm{S}^\mathrm{eff}(t) \ket{\psi(t)}
\end{equation}
can be written as
\begin{equation}
\ket{\psi(t)} = \sum_a d_a \ket{\psi_a(t)} = \sum_a d_a \, e^{- i E_a t}\ket{\phi_a(t)},
\end{equation}
where $\ket{\psi_a(t)} = e^{- i E_a t}\ket{\phi_a(t)}$ are the Floquet states with the \textit{periodic} Floquet modes $\ket{\phi_a(t+T)} = \ket{\phi_a(t)}$, quasi-energies $E_a$, and $d_a = \langle \phi_a(0) | \psi(0) \rangle$. By definition of the propagator~(\ref{propa}), we have
\begin{equation}\label{EVeq}
\ket{\psi_a(T)} = U(T) \ket{\psi_a(0)} 
\Leftrightarrow e^{- i E_a T}\ket{\phi_a(0)} = U(T) \ket{\phi_a(0)},
\end{equation}
showing that $e^{- i E_a T}$ are the eigenvalues of $U(T)$, which can be numerically computed using $U(T) \approx \prod_{n = 0}^{N} e^{ - i H_\mathrm{S}^\mathrm{eff}(n dt) dt}$ with $N = T/dt - 1$. Solving the eigenvalue problem~(\ref{EVeq}), we obtain $E_{a,k} = E_a + k \frac{2\pi}{T}$ with $k \in \mathbb{Z}$, and consider the values of $E_{a,k}$ lying in the first Brillouin zone $[-\pi/T, \pi/T]$ to define the quasienergies $E_a$. The eigenvectors correspond to the Floquet modes at initial time $\ket{\phi_a(0)}$. The Floquet modes at all times $t$ are obtained from these latter using
\begin{equation}
\ket{\phi_a(t)} = e^{i E_a t}U(t)\ket{\phi_a(0)}.
\end{equation}

\subsection{Master equation in the Floquet basis}

We now decompose the density matrix in the Floquet mode basis $\{\ket{\phi_a(0)}\}$, i.e.\ 
\begin{equation}\label{rhoI}
\rho(t) = \sum_{a,b} \rho^{I,ab}(t) \ket{\phi_a(0)} \bra{\phi_b(0)},
\end{equation}
and derive below the equations of motion for the density matrix element $\rho^{I,ab}(t) \equiv \bra{\phi_{a}(0)}\rho(t)\ket{\phi_b(0)}$ from Eq.~(\ref{EQ00}). Note that for the sake of clarity we restored the label $^I$ denoting the interaction picture for the density matrix elements.

In this basis, the matrix elements of the system operator $c_s(t)$ reads
\begin{equation}
\bra{\phi_{a}(0)}c_s(t)\ket{\phi_b(0)} = \bra{\phi_{a}(t)}c_s\ket{\phi_b(t)} e^{i (E_a - E_b) t}.
\end{equation}
Since $\ket{\phi_a(t)}$ is periodic of period $T$, we can rewrite $\bra{\phi_{a}(t)}c_s\ket{\phi_b(t)}$
in the Fourier space as
\begin{equation}\label{fourierseries}
\bra{\phi_{a}(t)}c_s\ket{\phi_b(t)} = \sum_{k\in \mathbb{Z}} e^{i k\Delta\mu t} c_s^{ab k},
\end{equation}
which yields
\begin{equation}
\bra{\phi_{a}(0)}c_s(t)\ket{\phi_b(0)} = \bra{\phi_{a}(t)}c_s\ket{\phi_b(t)} e^{i (E_a - E_b) t} = \sum_{k\in \mathbb{Z}} e^{i k \Delta_{ab k} t} c_s^{ab k} ,
\end{equation}
where $\Delta_{a b k} = E_a - E_b + k \Delta\mu$ and
\begin{equation}
c_s^{ab k} = \frac{1}{T} \int_0^T e^{-i k \Delta\mu t} \bra{\phi_{a}(t)}c_s\ket{\phi_b(t)} dt.
\end{equation}
Expanding all operators of the first term of the right-hand side of the master equation~(\ref{EQ00}) in the the Floquet basis yields
\begin{multline}
\langle\phi_a(0)| \left( \int_0^\infty dt' f_{\ell +}(t')  c_s(t)c_s^\dagger(t-t') \rho(t) \right) |\phi_b(0)\rangle
\\
= \sum_{c,d}\sum_{k,k'} e^{i (\Delta_{ack} + \Delta_{cdk'}) t} c_s^{ack} c_s^{\dagger cdk'} \rho^{I,db}(t) \int_0^\infty  dt' f_{\ell +}(t')  e^{-i \Delta_{cdk'} t'},
\end{multline}
and all other terms can be written in the same way. Hence, we see that the master equation involves complex rates
\begin{equation}\label{complexrates}
\begin{aligned}
\Gamma_{\ell\pm}(E) &= \int_0^\infty dt' f_{\ell \pm}(t') e^{i E t'} = \gamma_{\ell \pm}(E) + i \Omega_{\ell \pm}(E),
\end{aligned}
\end{equation}
where $E$ corresponds to system transition energies and where $\gamma_{\ell \pm}$  and $\Omega_{\ell \pm}$ are the real and imaginary parts of $\Gamma_{\ell \pm}$ which explicitly read
\begin{equation}\label{rates}
\begin{aligned}
&\gamma_{\ell \pm}(E) = \gamma_\ell  [1-n_\ell(E \pm \mu_\ell)], \\
&\Omega_{\ell \pm}(E) =  \frac{\gamma_\ell}{\pi} \,\mathrm{P.V.} \int_{-\infty}^{\infty} d\omega \frac{n_\ell(\omega)}{E + \omega \pm \mu_\ell},
\end{aligned}
\end{equation}
where $\mathrm{P.V.\ }$ denotes the principal value.
Note that the integrands appearing in the expressions of the shifts $\Omega_{\ell \pm}$ do not converge for $\omega \to \pm\infty$, and one has to introduce a cutoff frequency $\omega_c$ in the integration domain to obtain finite values for the shifts. In our simulations, we anyway neglected these shifts that are small compared to system energies -- since proportional to $\gamma_\ell$ -- and thus do not contribute significantly to the dynamics.

All together, the master equation~(\ref{EQ00}) written in the Floquet basis gives us the following set of equations for the matrix elements  $\rho^{I,ab}(t) \equiv \bra{\phi_{a}(0)}\rho^I(t)\ket{\phi_b(0)}$, i.e.\ the Floquet-Redfield master equation
\begin{equation}\label{EQF}
\begin{aligned} 
&\dot{\rho}^{I,ab}(t) = \sum_\ell (\mathcal{L}_\ell[\rho^I(t)])^{ab} \\
&= -\sum_{\ell } \sum_{s,k,k'}\sum_{c,d}\left\{ \left[ e^{i (\Delta_{cdk'} + \Delta_{ack})t}
 \left( c_s^{ack} c_s^{\dagger cdk'}  \Gamma_{\ell +}(-\Delta_{cdk'})  +
 c_s^{\dagger ack} c_s^{cdk'} \Gamma_{\ell -}(-\Delta_{cdk'})\right) \rho^{I,db}(t) \right. \right. \\
&\hspace{0.2cm}\phantom{\times\bigg[}
\left. \left. - e^{i (\Delta_{dbk'} + \Delta_{ack})t} \left(  c_s^{ack} c_s^{\dagger dbk'}  \Gamma_{\ell -}(-\Delta_{ack}) + c_s^{\dagger ack} c_s^{dbk'} \Gamma_{\ell +}(-\Delta_{ack})\right) \rho^{I,cd}(t) \right] + \mathrm{h.c.} \right\}.
\end{aligned}
\end{equation}
The master equation for the matrix elements in Schrödinger picture $\rho^{ab}(t)$ can be obtained from Eq.~(\ref{EQF}) by making the replacement $\rho^{I,ab}(t) = e^{i (E_a - E_b)t} \rho^{ab}(t)$. In doing so, one can see that all terms $e^{i(\Delta_{cdk'} + \Delta_{ack})t}$ reduces to $e^{i(k + k')\Delta\mu t}$ [see Eq.(\ref{EQF0})], showing that the master equation exhibits the same periodicity than the system Hamiltonian $H_\mathrm{S}^\mathrm{eff}(t)$. This implies that the steady state of the master equation is also periodic with the same period $T = 2\pi/\Delta\mu$~\cite{Yud16}.

Note that we did not proceed with the secular approximation, so that we have a Redfield-like master equation, for which the steady state properties match the ones of the equivalent (weak-coupling) non-Markovian master equation for time-independent system. Hence, while non-Markovian effects might potentially be present in the transient dynamics of our system, we do not expect significant memory effects in its steady state properties. The investigation of the interplay between potential non-Markovian effects and Floquet dynamics will be investigated in a further work.

\section{Solution of the master equation and particle currents}

We present here two methods to solve the master equation~(\ref{EQF}) that exploits its periodicity. Note that since the steady state density matrix and any expectation value of system operators obtained from it are in principle periodic (or constant), we always present in the main text time-averaged values of these quantities over one period of oscillation $T = 2\pi/\Delta \mu$.

The master equation in Schrödinger picture~(\ref{EQF0}) can be vectorized in the form
\begin{equation}\label{vecME}
\ket{\dot{\rho}^S(t)} = L(t)  \ket{\rho^S(t)}
\end{equation}
where $\ket{\rho^S(t)} = (\rho^{S,11}(t),\rho^{S,12}(t),\dotsc, \rho^{S,44}(t)) $ is the vectorized density matrix and $L(t)$ is a periodic time-dependent matrix of period $T$. 

\subsection{Solving the master equation in Fourier space}

We can express the steady state density matrix $\rho_{SS}^S(t)$ and the matrix $L(t)$ as
\begin{equation}
\begin{aligned}
&\rho_{SS}^S(t) = \sum_{k} e^{-i k \Delta\mu t} \rho_k\mbox{;} \hspace{0.5cm}
\rho_k = \frac{1}{T} \int_0^T e^{-i k \Delta\mu t} \rho_{SS}^S(t) dt \\
&L(t) = \sum_{k} e^{-i k \Delta\mu t} L_{k} \mbox{;} \hspace{0.5cm}
L_{k} = \frac{1}{T} \int_0^T e^{-i k \Delta\mu t} L(t) dt.
\end{aligned}
\end{equation}
Inserting these decompositions into the vectorized form~(\ref{vecME}), we get
\begin{equation}
\sum_k -i k V e^{-i k \Delta\mu t} \ket{\rho_k} = \sum_{k, k'} e^{-i (k + k') \Delta\mu t} L_{k'}\ket{\rho_k}
\end{equation}
Applying then $(1/T)\int_0^T \;\; \cdot\;\; e^{i k'' \Delta\mu t}$ on both sides yields
\begin{equation}
0 = -i k'' V \mathbb{1} \ket{\rho_{k''}} + \sum_k L_k \ket{\rho_{k''-k}}
\end{equation}
which can be written in matrix form as
\begin{equation}\label{MEinFS}
\begin{pmatrix} 
\vdots \\
0 \\
0 \\
0 \\
\vdots 
\end{pmatrix} =
\begin{pmatrix}
\ddots  & & & & \\
 & -i (k + 1) \Delta\mu  + L_0 & L_1 & L_2 &  \\
 & L_{-1} & -i k \Delta\mu  + L_0 & L_1 &  \\
 & L_{-2} & L_{-1} & -i (k-1) \Delta\mu  + L_0 & \\
 & & & &  \ddots \\
\end{pmatrix}
\begin{pmatrix}
\vdots \\
\ket{\rho_{k+1}} \\
\ket{\rho_{k}}  \\
\ket{\rho_{k-1}}  \\
\vdots
\end{pmatrix}
\end{equation}
This linear system of equations for the Fourier components can be efficiently solved numerically after truncation, i.e., by introducing a cutoff $k_\mathrm{max}$ in the summation over $k$ of the Fourier series~(\ref{fourierseries}). The off-diagonal blocks $L_k$ in Eq.~(\ref{MEinFS}) describe the coupling between different Fourier components. A block $L_k$ corresponds to $k$ molecular conversion process of the form~(\ref{conversion}).

\subsection{Solving the master equation in real space}

The vectorized master equation~(\ref{vecME}) is of the same form than the time-dependent Schrodinger equation~(\ref{TDse}). We can thus apply again the Floquet theory and write the solution as
\begin{equation}
\ket{\rho^S(t)} = \sum_{a} d_a e^{ \mu_\alpha t} \ket{\rho^{S,a}(t)}
\end{equation}
where $\ket{\rho^{S,a}(t)} = \ket{\rho^{S,a}(t+T)}$ are periodic functions of period $T$ and where $d_\alpha = \langle \rho^{S,a}(0) | \rho^S(0) \rangle $. Hence, solving the entire problem in this case consists in applying twice the Floquet theory: once to write the Floquet-Redfield master equation~(\ref{EQF}) and once to solve it.

\subsection{Currents}

From the solution of the master equation~(\ref{EQF}) for the matrix elements $\rho^{I,ab}(t)$, one can evaluate the expectation values of any system operator $O$ through
\begin{equation}\label{expop}
\begin{aligned}
\langle O \rangle &= \mathrm{Tr}\left[ U^\dagger(t) O U(t) \rho^I(t) \right] = \sum_{j = 0,\downarrow,\uparrow,\downarrow \uparrow}\sum_{ab} \rho^{I,ab}(t) e^{-i (E_a - E_b) t} \bra j O \ket{\phi_a(t)} \bra{\phi_b(t)} j \rangle.
\end{aligned}
\end{equation}
We derive here the expressions of the particle currents $I_\mathrm{S}$, $I_\ell$ and $I_{\mathrm{mol}}$ respectively in the junction, the reservoirs and the molecular BEC. For the sake of clarity, we introduce an annihilation operator $S$ for the BEC that interacts with the junction with an Hamiltonian of the form $H_\mathrm{BEC}(t) = \sum_\ell ( \Omega_\ell S c^\dagger_\uparrow c^\dagger_\downarrow e^{2 i \mu_\ell t}+ \mathrm{h.c.} )$. The currents $I_\mathrm{S}$, $I_\ell$ and $I_{\mathrm{mol}}$ are then defined as
\begin{align}
&I_\mathrm{S} = \frac{d}{dt} \sum_{s} c_{ s}^\dagger c_{s} = -i \kappa_\ell \sum_{ks} \left( c_s^\dagger b_{\ell k s } - \mathrm{h.c.}\right) - 2 i   \sum_\ell \Big(  g_\ell^* e^{-2i\mu_\ell t} c_\uparrow^\dagger c_\downarrow^\dagger - \mathrm{h.c.} \Big), \\
&I_\ell = \frac{d}{dt} \sum_{ks} b_{\ell k s}^\dagger b_{\ell k s} = i \kappa_\ell \sum_{ks} \left( c_s^\dagger b_{\ell k s } - \mathrm{h.c.}\right), \\
&I_{\mathrm{mol}} = \frac{d}{dt} S^\dagger S = i \sum_\ell \left( g_\ell^* e^{-2i \mu_\ell t} c_\uparrow^\dagger c_\downarrow^\dagger - \mathrm{h.c.} \right), \label{Ip}
\end{align}
where we replaced the operator $S$ by the macroscopic fraction $\langle S\rangle$  and where we used the Langevin equations for $c_s$, $b_{\ell k s}$ and $S$. Note that the total number of particles is conserved, i.e.,
\begin{equation}\label{continuity}
I_\mathrm{S} + \sum_{\ell} I_\ell + 2 I_\mathrm{mol} = 0,
\end{equation}
where the factor $2$ is front of $I_\mathrm{mol}$ denotes the fact that a molecule is made of two atoms.

The expectation value of the particles current in the junction is obtained from the solutions of the master equation and Eq.~(\ref{expop}), that is
\begin{equation}
\begin{aligned}
\langle I_\mathrm{S} \rangle = \sum_s \frac{d}{dt}\langle c_s^\dagger c_s \rangle  &= \sum_s \frac{d}{dt}\mathrm{Tr}\left[U^\dagger(t) c_s^\dagger c_s U(t) \rho^I(t) \right].
\end{aligned}
\end{equation}
Due to the conservation of the number of particles~(\ref{continuity}), it can be related to the particle current in the reservoirs, as shown below. Applying the derivative and using the fact that $dU(t)/dt = -i H_\mathrm{S}^\mathrm{eff}(t) U(t)$, we get
\begin{equation}\label{IQDtemp}
\begin{aligned}
\langle I_\mathrm{S} \rangle
&= i \sum_s \mathrm{Tr}\left[ U^\dagger(t) H_\mathrm{S}^\mathrm{eff}(t) c_s^\dagger c_s U(t)\rho^I(t) \right]  - i  \sum_s \mathrm{Tr}\left[ U^\dagger(t) c_s^\dagger c_s H_\mathrm{S}^\mathrm{eff}(t) U(t) \rho^I(t)  \right] \\
&\hspace{2cm} + \sum_s \mathrm{Tr}\left[ U^\dagger(t) c_s^\dagger c_s U(t) \frac{d\rho^I(t)}{dt}  \right]. 
\end{aligned}
\end{equation}
The two first terms on the right-hand-side can be rewritten as
\begin{equation}
\begin{aligned}
i \sum_s \langle H_\mathrm{S}^\mathrm{eff}(t) c_s^\dagger c_s \rangle  - i  \sum_s \langle c_s^\dagger c_s  H_\mathrm{S}^\mathrm{eff}(t) \rangle &= 2 i \sum_\ell g_\ell e^{2i \mu_\ell t} \langle c_\downarrow c_\uparrow \rangle - 2 i \sum_\ell g_\ell^* e^{-2i \mu_\ell t} \langle c_\uparrow^\dagger c_\downarrow^\dagger \rangle \\
&= - 2 \langle I_\mathrm{mol} \rangle
\end{aligned}
\end{equation}
where [see Eq.~(\ref{Ip})]
\begin{equation}\label{Ipl}
\langle I_\mathrm{mol} \rangle = i \sum_\ell \left( g_\ell^* e^{-2i \mu_\ell t} \langle c_\uparrow^\dagger c_\downarrow^\dagger \rangle - \mathrm{h.c.} \right).
\end{equation}
Finally, replacing the derivative in Eq.~(\ref{IQDtemp}) by the right-hand-side of the master equation~(\ref{EQF}) yields
\begin{equation}
\begin{aligned}
\langle I_\mathrm{S} \rangle 
&= - 2 \langle I_\mathrm{mol} \rangle +  \sum_{\ell s} \mathrm{Tr}\left[ U^\dagger(t) c_s^\dagger c_s U(t) \left(\mathcal{L}_\ell \left[  \rho^I(t)\right]\right) \right] = - 2  \langle I_\mathrm{mol} \rangle - \sum_\ell \langle I_\ell \rangle ,
\end{aligned} 
\end{equation}
with
\begin{equation}\label{Isl}
\begin{aligned}
\langle I_\ell \rangle &= - \sum_{ab} \left(\mathcal{L}_\ell\left[\rho^I(t) \right]\right)^{ab}  e^{-i (E_a - E_b)t}  \bigg( \sum_{j}\sum_s \langle j | c_s^\dagger c_s  \ket{\phi_a(t)} \bra{\phi_b(t)}  j \rangle \bigg).
\end{aligned}
\end{equation}
In the main text, we always present time-averaged values of the current over one period of oscillation $T = 2\pi/\Delta \mu$.

\subsection{Numerical details}

In this section, we provide the numerical details of the resolution of the master equation~(\ref{EQF0}). First, in order to write the master equation, we computed the quasi-energies from Eq.~(\ref{EVeq}) using the procedure stated above with $N = 1000$, a number of time steps which insures a sufficient convergence of the quasi-energies to compute the particle currents. Writing the master equation also requires to define a cutoff $k_\mathrm{max}$ of the Fourier series. We define $k_\mathrm{max}$ as an empirical function of $U$ and $\Delta\mu$ also so that the computed current has converged up to a given precision. Typical values range from $k_\mathrm{max} \sim 4$ for large bias and interaction to $k_\mathrm{max} \sim 80$ for small bias and interaction. To solve the master equation~(\ref{EQF0}), we use brute force resolution of the differential equations, with random or particular initial states, from a initial time $t_i = 0$ to a final time $t_f = 10 \gamma^{-1}$, where $\gamma \equiv \gamma_\ell = 10^{-2}$ (Fig. 2 and 4) or $10^{-3}$ (Fig. 3), in order to reach the steady state. We then averaged the particle currents over one period $T$. All together, we required that any increase of the precision (which can be achieved via the parameters $N$, $k_\mathrm{max}$ and $t_f$) yields only an improvement of the value of $\langle I_R \rangle/\gamma$ smaller than $10^{-2}$. We also checked that the alternative methods proposed in Sec. B.1 and B.2 to solve the master equation give similar results.

\section{Adding dissipation on the channel}

Additional Lindblad dissipation acting on the channel can be accounted for by adding to Eq.~(\ref{EQF}) a dissipator of the form
\begin{equation}
\begin{aligned}
\mathcal{D}_\mathrm{I}[\rho^S(t)] &= \gamma_\mathrm{I} \bigg(2 L \rho^S(t) L^\dagger 
- L^\dagger L \rho^S(t)
-  \rho^S(t) L^\dagger L \bigg)
\end{aligned}
\end{equation}
In the Floquet basis and in interaction picture with respect to $H_\mathrm{S}^\mathrm{eff}(t)$, this dissipator reads
\begin{equation}\label{Dincoh}
\begin{aligned}
(\mathcal{D}_\mathrm{I}[\rho^I(t)])^{ab}
&= \gamma_\mathrm{I}\sum_{cd}\sum_{kk'} \bigg[ 
2\left( e^{i (\Delta_{dbk'} + \Delta_{ack})t} L^{ack} L^{\dagger dbk'}\right) \rho^{I,cd}(t) \\
&\hspace{0.7cm}
- \left(  e^{i (\Delta_{cdk'} + \Delta_{ack})t} L^{\dagger ack} L^{cdk'} \right) \rho^{I,db}(t) 
- \left(  e^{i (\Delta_{dbk'} + \Delta_{cdk})t} L^{\dagger cdk} L^{dbk'}\right) \rho^{I,ac}(t) \bigg].
\end{aligned}
\end{equation}
The associated particle current reads
\begin{equation}\label{IIncoh}
\langle I_\mathrm{I}(t) \rangle =-\sum_{a,b} \left(\mathcal{D}_\mathrm{I}\left[\rho^I(t) \right]\right)^{ab}  e^{-i (E_a - E_b)t} \bigg( \sum_{j}\sum_s \langle j | c_s^\dagger c_s  \ket{\phi_a(t)} \bra{\phi_b(t)}  j \rangle \bigg). \\
\end{equation}

\bibliographystyle{unsrt}
\bibliography{mybib}

\begin{thebibliography}{10}
\bibitem{CommentNJP2019001}
In cold atom experiments, the initial imbalance between the reservoirs is
  usually not maintained during the transport measurements. This is in contrast
  with the solid-state where the bias between the leads that act as reservoirs
  can easily be conserved through the use of bias voltages.
\bibitem{C2}
The condition $\tau_R \gg \tau_B$ can also be fulfilled for single-mode
  environment that are damped by other means, as it is the case for an atomic
  system trapped in a single-mode lossy cavity~\cite{Dam19a}.
\bibitem{C3}
Note that we observed a slower numerical convergence of the transport
  properties of the junction when decreasing the interaction strength $|U|$, in
  the sense that a higher cutoff $k_\mathrm{max}$ of the Fourier series
  appearing in~(\ref{Liouvillian}) was required in this regime. This can be
  understood by the fact that for smaller interaction $|U|$ compared to a fixed
  pair-tunnelling $g$, higher-order multiple Andreev reflections must be
  accounted for, requiring then higher $k_\mathrm{max}$.
\bibitem{ComLoss}
We emphasise again that these effects can be directly modelled within the
  present framework, as done for the few examples presented in Sec.~3.4.
\end{thebibliography}

\end{document}